\documentclass[a4paper,12pt]{article}

\def\beq{\begin{equation}}
\def\eeq{\end{equation}}
\def\beqn{\begin{eqnarray}}
\def\eeqn{\end{eqnarray}}
\def\nn{\nonumber\\ }

\usepackage{cite}
\usepackage{wrapfig}
\usepackage{bm}
\usepackage{amssymb}
\usepackage{times}
\usepackage{mathptmx}
\usepackage[dvips]{graphicx}

\usepackage[a4paper,hmargin={30mm,15mm},vmargin={25mm,25mm}]{geometry}

\begin{document}

\begin{center}
ON EXTRACTION OF THE TOTAL PHOTOABSORPTION CROSS SECTION ON THE NEUTRON
FROM DATA ON THE DEUTERON
\\[1ex]
M.I.~Levchuk$^1$, A.I.~L'vov$^2$
\\[1ex]
{$^1$\it B.I. Stepanov Institute of Physics of the National Academy of Sciences of Belarus, Minsk}\\
{$^2$\it P.N. Lebedev Physical Institute of the Russian Academy of Sciences, Moscow, Russia}\\

\vspace{3ex}

\begin{minipage}{0.9\textwidth}
\small
An improved procedure is suggested for finding the total photoabsorption
cross section on the neutron from data on the deuteron at energies
$\lesssim 1.5$ GeV. It includes unfolding of smearing effects caused by
Fermi motion of nucleons in the deuteron and also takes into account
non-additive contributions to the deuteron cross section due to 
final-state interactions of particles in single and double pion
photoproduction. This procedure is applied to analysis of existing data.
\end{minipage}

\end{center}

\subsection*{Introduction}

This work was motivated by recent preliminary results from the GRAAL
experiment on the total photoabsorption cross section off protons and
deuterons at photon energies $\omega=700{-}1500$ MeV
\cite{bartalini08,rudnev10,nedorezov12,turinge12} and their implications
for the neutron.
An intriguing feature of the new data is that they indicate an
approximately equal and big strength of photoexcitation of the nucleon
$F_{15}(1680)$ resonance off both the proton and neutron (as seen, in
particular, in Fig.~5 in Ref.~\cite{turinge12}).
Meanwhile this strength was found small for the neutron in many previous
studies (see, e.g., \cite{armstrong72a,armstrong72b}). Particle Data
Group \cite{PDG12} quotes the following branching ratios of $N^* =
F_{15}(1680)$ to $\gamma N$:
\beqn
       {\rm Br}(N^* \to \gamma p) &=& 0.21{-}0.32\%,
\nn
       {\rm Br}(N^* \to \gamma n) &=& 0.021{-}0.046\%.
\eeqn
Irrespectively on whether the old or new data are correct, it seems
timely to (re)consider procedure commonly used to find cross sections
off the neutron from the deuteron data.

This procedure was described in detail by the Daresbury group
\cite{armstrong72b} who performed measurements of the total
photoabsorption cross sections $\sigma_p$ \cite{armstrong72a} and
$\sigma_d$ \cite{armstrong72b} at energies between 0.265 and 4.215 GeV.
In the nucleon resonance energy region they made an Ansats that 
\beq
     \sigma_d(\omega) = F(\omega)[\sigma_p(\omega) + \sigma_n(\omega)].
\label{armstrong-ansatz}
\eeq
Here the factor of $F(\omega)$ was introduced in order to take into
account smearing effects due to Fermi motion of nucleons in the
deuteron. This factor was found by numerical integration of the proton
cross sections using known momentum distribution of nucleons in the deuteron and
then equally applied to the neutron. Finally, the neutron cross section
was found, point by point, with the step of 25 MeV, from the corresponding deuteron cross section
at the same energy using Eq.~(\ref{armstrong-ansatz}).

An evident drawback of the Ansatz (\ref{armstrong-ansatz}) is that
smearing effects are assumed to be the same for the proton and neutron,
what cannot be true in case the energy dependencies of
$\sigma_p(\omega)$ and $\sigma_n(\omega)$ are different. 

The second problem is that smearing of the cross section makes it
impossible to relate individual nucleon cross sections
$\sigma_N(\omega)$ with $\sigma_d(\omega)$ at the same energy and thus
to apply the point-by-point procedure. Instead, some average of
$\sigma_N(\omega)$ over a finite energy interval can only be found. In
other words, a justified unsmearing procedure should be applied there.

The third point is that non-additive corrections related mostly with
final state interactions have been neglected in
Eq.~(\ref{armstrong-ansatz}). Brodsky and Pumplin \cite{brodsky69}
estimated these corrections at high energies ($\omega\gtrsim 2$ GeV)
assuming that high-energy photoproduction on the nucleon is dominated by
diffractive photoproduction of vector mesons ($\rho$, $\omega$, $\phi$)
which then interact with the second nucleon. Such corrections have been
included in the analysis of high-energy part of the Daresbury data
\cite{armstrong72b} (as well as in studies of photoabsorption off
protons and deuterons at energies 20--40 GeV \cite{belousov75}). At
lower energies, including energies of GRAAL, the corrections related
with vector meson production are small. Nevertheless, other
photoproduction channels still might be important. This is indeed the
case as explained below. To our knowledge, no estimates of the
non-additive corrections to Eq.~(\ref{armstrong-ansatz}) have been yet
done at energies of the GRAAL experiment.

In this work we improve the procedure of \cite{armstrong72b} in all the
above three lines.

\subsection*{Fermi smearing (folding)}

\begin{figure}[htb]
\centering\includegraphics[width=0.5\textwidth]{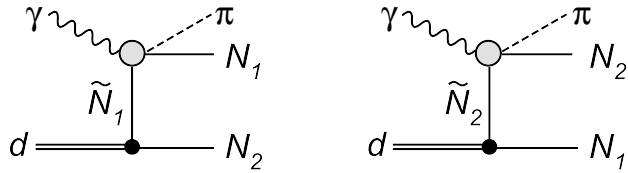}
\caption{Diagrams of impulse approximation for $\gamma d\to\pi NN$.
Antisymmetrization over $N_1$ and $N_2$ is not shown.}
\label{diagIA}
\end{figure}

We begin with rewriting Eq.~(\ref{armstrong-ansatz}) more accurately as
\beq
     \sigma_d(\omega) = \hat F 
        [\sigma_p(\omega) + \sigma_n(\omega)] + \Delta\sigma_{pn}(\omega).
\label{sigma-d}
\eeq
Here $\hat F$ is a linear integral operator that smears individual
nucleon cross sections in accordance with Fermi motion of nucleons in
the deuteron; $\Delta\sigma_{pn}$ is a non-additive correction to be
discussed later. The first two terms in Eq.~(\ref{sigma-d}) arise from
diagrams of impulse approximation (like those in Fig.~\ref{diagIA}) when
interference effects are omitted. We neglect here off-shell effects for
intermediate nucleons $\tilde N$ because the binding energy of nucleons
in the deuteron is rather small (2.2 MeV).

\begin{figure}[htb]
\centering\includegraphics[width=0.35\textwidth]{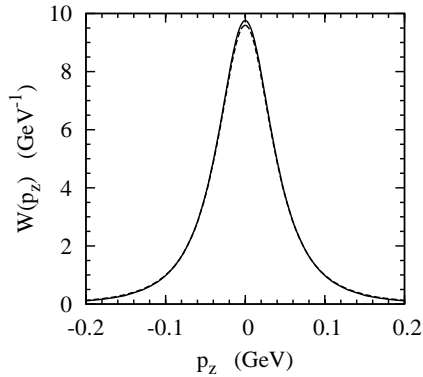}
\vspace{-1em}
\caption{Distribution of the longitudinal momentum in the deuteron.
Solid and dashed lines: Hulth\'en and CD-Bonn wave functions.}
\vspace{1em}
\label{Wpz}
\end{figure}

A simple analysis of diagrams of impulse approximation shows
\cite{west72} that the smearing operator, in nonrelativistic
approximation over nucleons in the deuteron, is reduced to
\beq
    \hat F\sigma_N(\omega) = \int W(p_z)\frac{\omega^{\rm eff}}{\omega}
        \,\sigma_N(\omega^{\rm eff})\,d\!p_z.
\label{F}
\eeq
Here 
\beq
   \omega^{\rm eff} = \omega \Big(1-\frac{p_z}{M}\Big)
\eeq
is the effective (Doppler shifted) energy for the moving intermediate nucleon
$\tilde N$ of the mass $M$ provided its longitudinal (along the photon
beam) momentum is equal to $p_z$. $W(p_z)$ is the longitudinal momentum distribution
of nucleons in the deuteron,
\beq
    W(p_z) = \int |\psi(p)|^2 \frac{d^2p_\perp}{(2\pi)^3},
\eeq
and the factor $\omega^{\rm eff}/\omega$ takes into account a change in
the photon flux seen by the moving nucleon.
As in Ref.~\cite{armstrong72b}, we use in the following a simplified deuteron
wave function (Hulth\'en \cite{hulthen57}),
\beq
   \psi(r) = \frac{k}{r}(e^{-ar} - e^{-br}), \qquad \int_0^\infty |\psi(r)|^2 \,4\pi r^2 d\!r =1,
\eeq
with $a=45.7~{\rm MeV}\!/c$, $b=260~{\rm MeV}\!/c$ and
$k^2 =  ab(a+b)/[2\pi(a-b)^2] = 12.588~{\rm MeV}\!/c$.
In the $p$-space
\beq
   \psi(p) = 4\pi k \Big( \frac{1}{a^2 + p^2} - \frac{1}{b^2 + p^2} \Big),
\eeq
so that the function $W(p_z)$ is
\beq
   W(p_z) = 2k^2 \Big( \frac{1}{A} + \frac{1}{B} - \frac{2\ln(B/A)}{B-A} \Big),
   \qquad \int W(p_z)\,d\!p_z=1,
\label{W-p_z}
\eeq
where $A = a^2+p_z^2$ and $B = b^2+p_z^2$. This function is shown in
Fig.~\ref{Wpz} together with a distribution obtained with a realistic
(CD-Bonn) wave function \cite{CD-Bonn}. In actual calculations we cut
off momenta $|p_z| > p_{\rm cut} = 200~{\rm MeV}\!/c$ where $W(p_z)$
becomes quite small and the momentum $p_z$ remains nonrelativistic.

The Hulth\'en distribution for $W(p_z)$ gives the following average longitudinal
momentum of nucleons in the deuteron:
\beq
    \langle p_z^2 \rangle^{1/2} = 53.9~{\rm MeV}\!/c
\eeq
(it is $54.9~{\rm MeV}\!/c$ for the CD-Bonn wave function).
It also gives the following spread for the effective photon energy seen
by the moving nucleon:
\beq
    \Delta\omega^{\rm eff} = \omega \frac{\langle p_z^2 \rangle^{1/2}}{M} = 0.057\omega.
\eeq
In other words, this value characterizes the ``energy resolution of the
deuteron'' as a ``spectral measuring device'' for the neutron. For
$\omega \sim 1$ GeV only an average of the nucleon cross section over
the range $\sim\pm 60$ MeV can be inferred from the deuteron data.
Determination of $\sigma_n(\omega)$ with the step of 25 MeV done in
\cite{armstrong72b} cannot be physically justified.

\subsection*{Unfolding}

It is well known that the unfolding problem, i.e. solving the Fredholm
integral equation (\ref{sigma-d}) for the unknown ``unsmeared deuteron
cross section'' $\sigma(\omega) = \sigma_p(\omega) + \sigma_n(\omega)$,
cannot be solved without further assumptions on properties of the
solution $\sigma(\omega)$. In particular, it is not possible to restore
fast fluctuations in $\sigma(\omega)$ at the energy scale
$\lesssim\Delta\omega^{\rm eff}$. To proceed, we make therefore a
physically sound assumption that both the cross sections
$\sigma_p(\omega)$ and $\sigma_n(\omega)$ can be approximated with a sum
of a few Breit-Wigner resonances (having fixed known standard masses and
widths but unknown amplitudes, probably different for $p$ and $n$) plus a
smooth background. Thus we write
\beq
     \sigma(\omega) = \sum_i X_i \,f_i(\omega)
\label{sX}
\eeq
where $f_i(\omega)$ is the basis of the expansion, i.e. either
Breit-Wigner distributions or smooth functions of the total energy
$\sqrt s$. We borrow specific forms of the functions $f_i(\omega)$ from
Ref.~\cite{armstrong72b}, Eqs.~(11) and below. Then unknown coefficients
$X_i$ are determined from the fit of $\hat F\sigma(\omega)$ to
experimental data on $\sigma_d(\omega)$ (at this point we assume that
the correction $\Delta\sigma_{pn}$ is already calculated).

A knowledge of $X_i$, with errorbars $\delta X_i$ determined in the fit,
can be directly converted to the knowledge of $\sigma(\omega)$, also
with errorbars. In particular, writing fluctuations in the determined
value of $\sigma(\omega)$ as
\beq
     \delta\sigma(\omega) = \sum_i \delta X_i \,f_i(\omega),
\eeq
we have 
\beq
     \delta\sigma^2(\omega) = \sum_{ij} \delta X_i \,\delta X_j \,f_i(\omega) f_j(\omega)
\eeq
and 
\beq
     \langle\delta\sigma^2(\omega)\rangle = \sum_{ij} C_{ij} \,f_i(\omega) f_j(\omega),
\label{err-s}
\eeq
where 
\beq
     C_{ij} = \langle \delta X_i \,\delta X_j\, \rangle
\eeq
is a standard covariance matrix of errors determined in the fit of $X_i$.

In this way the extracted unfolded cross section $\sigma(\omega)$ can be
shown as a smooth curve (corresponding to the central values of
$X_i$) surrounded with a band of the half-width given by Eq.~(\ref{err-s}) which 
represents errors in the cross section.

\subsection*{Nonadditive corrections}

The term $\Delta\sigma_{pn}(\omega)$ in Eq.~(\ref{sigma-d}) takes into account various effects
violating additivity of the photoabsorption cross sections on individual nucleons. Among them:

-- interference of diagrams of photoproduction off proton and neutron,
Fig.~\ref{diagIA}, leading to identical final states; the Fermi
statistics of the emitted nucleons (antisymmetrization) leading to the
so-called Pauli blocking,

-- interaction between emitted particles (final state interaction, FSI) including both interaction
of unbound nucleons and binding of nucleons (formation of the deuteron in the final state), interaction of pions
(or other particles), produced on one nucleon, with the second nucleon in the deuteron,

-- absorption of pions (and the presence of processes such as the deuteron photodisintegration,
without pions in the final state).

\begin{figure}[htb]
\centering\includegraphics[width=0.7\textwidth]{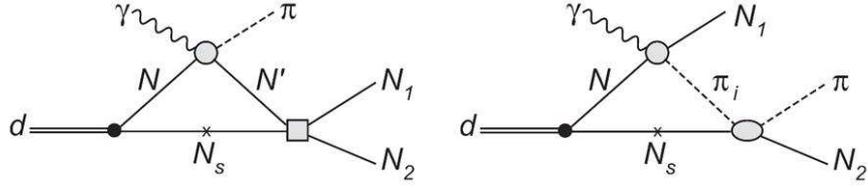}
\caption{Diagrams with the final state $NN$ and $\pi N$ interaction (to one loop) for $\gamma d\to\pi NN$.}
\label{diagFSI}
\end{figure}

\begin{figure}[htb]
\vspace{2em}
\centering
\includegraphics[width=0.40\textwidth]{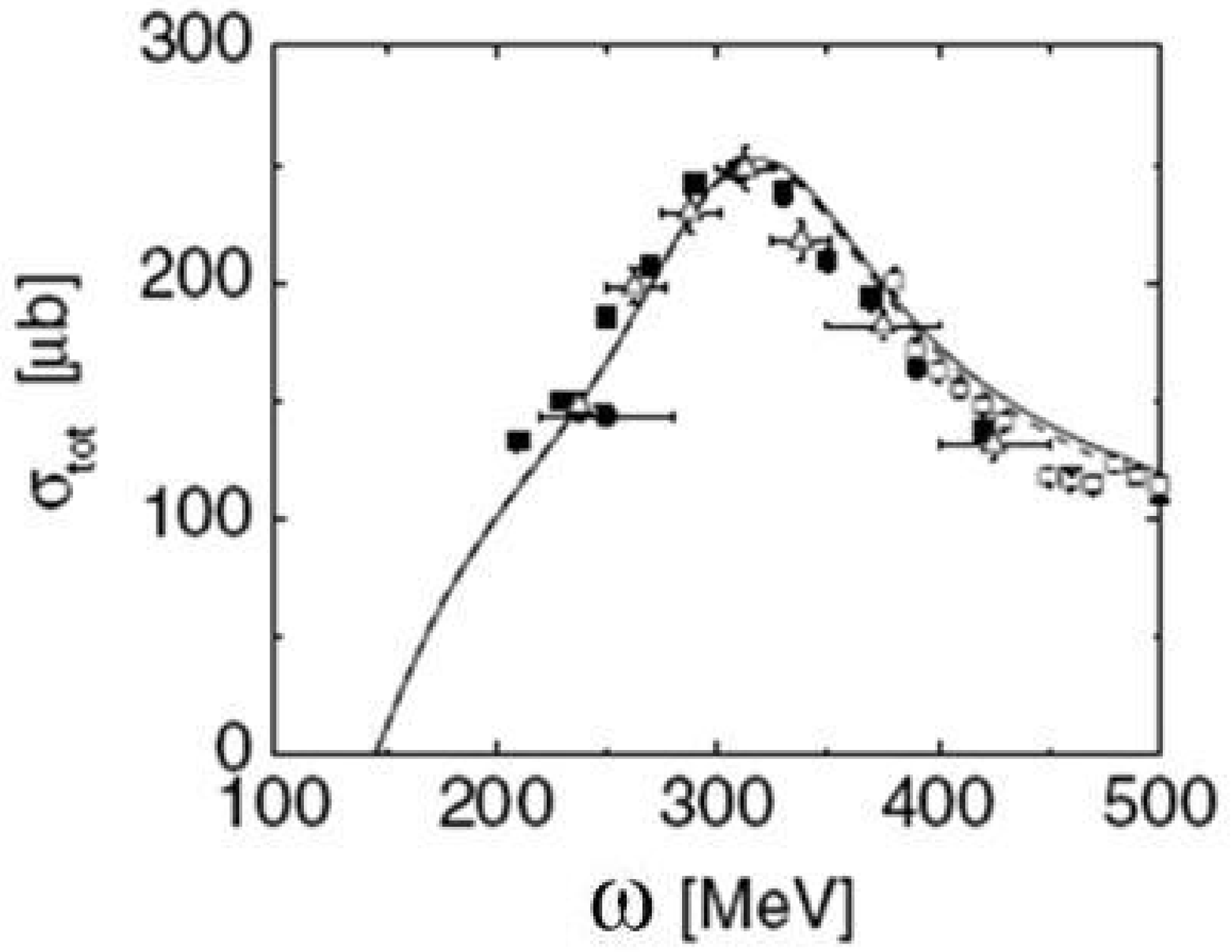}
\includegraphics[width=0.40\textwidth]{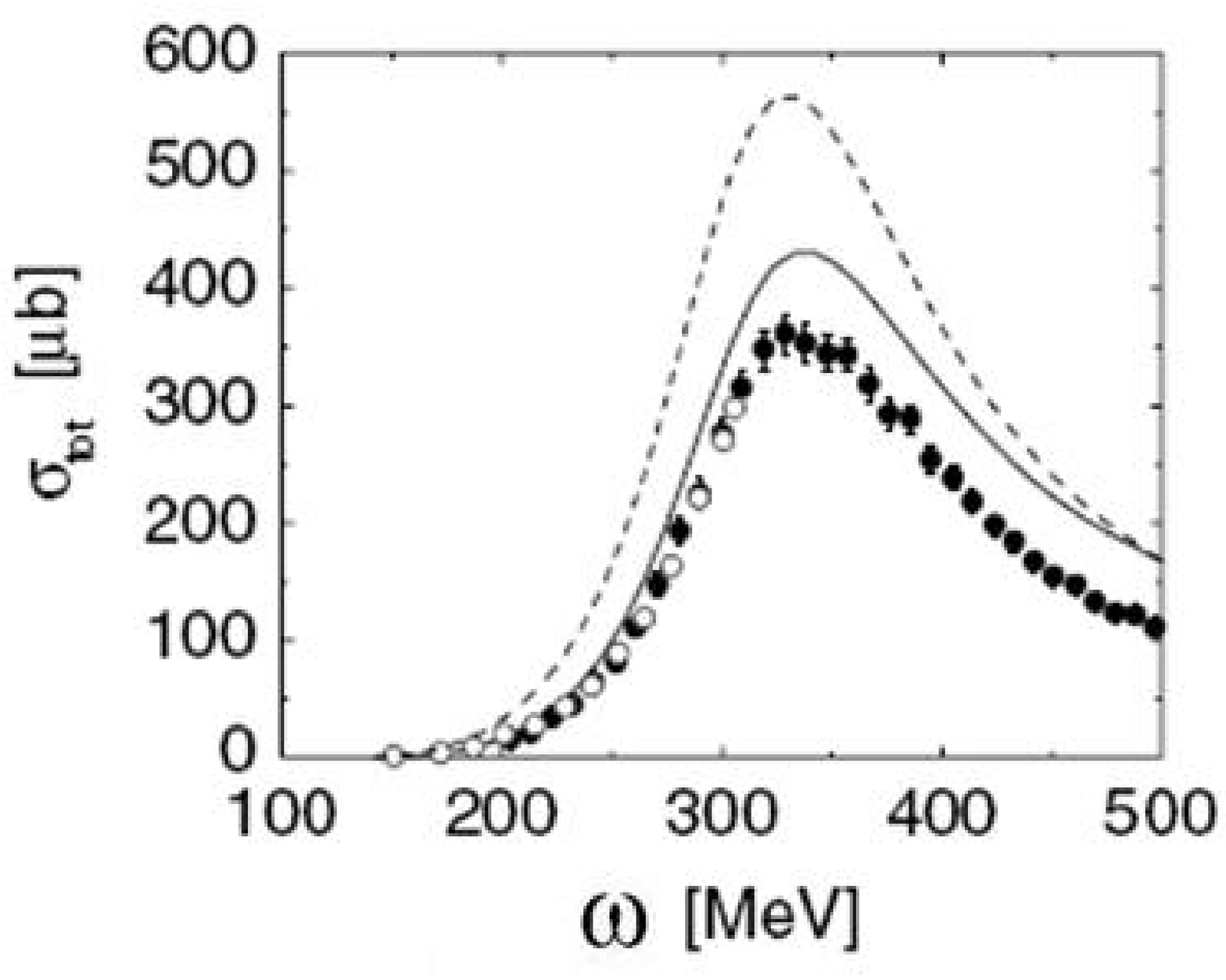}
\caption{Model \cite{levchuk06,levchuk10} predictions for $\gamma
d\to\pi^-pp$ (left) and $\gamma d\to\pi^0pn$ (right) in the region of
the $\Delta(1232)$.}
\label{cs.piNN}
\end{figure}

Now we briefly discuss all these effects starting with the reaction of
single-pion photoproduction, $\gamma d\to\pi NN$, considered in
the model that includes diagrams of impulse approximation
(Fig.~\ref{diagIA}) and the final state $NN$ and $\pi N$ interaction to one
loop (Fig.~\ref{diagFSI}). Formalism and the main building blocks of
this model that was previously used in the energy region of the
$\Delta(1232)$ resonance can be found elsewhere
\cite{levchuk06,levchuk10}. Generally, the model works well for the
channel $\gamma d\to\pi^-pp$ in the $\Delta(1232)$ region but not so well for
$\gamma d\to\pi^0pn$, see Fig.~\ref{cs.piNN}. Reasons for the
discrepancy are not clear but other authors get similar results and also
cannot describe the data (see, e.g., \cite{schwamb10}). We will not use
the model for energies too close to the $\Delta(1232)$ region.

In the present calculation that covers higher energies, ``elementary''
amplitudes of $\gamma N\to\pi N$ are taken from the MAID analysis
\cite{MAID} (with a proper off-shell extrapolation); those for $NN\to
NN$ are taken from the analysis of SAID \cite{SAID} (again with an
off-shell extrapolation). In the following plots we show obtained
results for $\Delta\sigma_{pn}(\omega)$ in different isotopic channels.

\subsubsection*{1. Interference contributions from diagrams of impulse approximation for $\bm{\gamma d\to\pi N\!\!N}$}

\begin{figure}[htb]
\centering\includegraphics[width=0.48\textwidth]{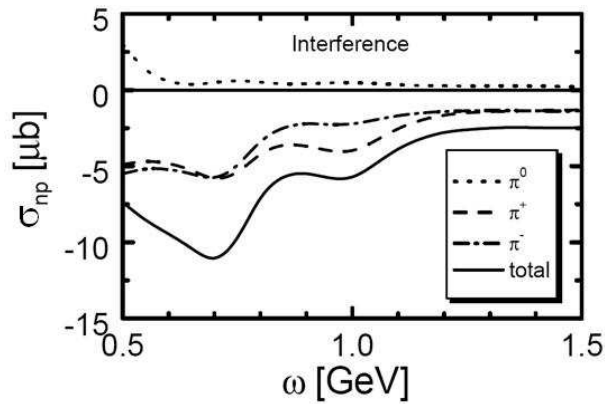}
\caption{Contribution to $\Delta\sigma_{pn}$ due to interference of diagrams in Fig.~\ref{diagIA} of impulse approximation
 for $\gamma d\to\pi NN$.}
\label{deltaS.IA}
\end{figure}

\subsubsection*{2. $\bm{N\!\!N}$ FSI interaction in $\bm{\gamma d\to\pi N\!\!N}$ and $\bm{\gamma d\to\pi d}$}

We put here $NN$ FSI contributions for the continuous and bound states
together because there is a tendency for their cancelation that can be
traced to the unitarity (closure). The matter is that the $NN$
interaction in the continuous spectrum can be thought as a
replacement of the plane $NN$ wave in the reaction amplitude of the
plane-wave impulse approximation,
\beq
   T^{\rm PWIA}(E_{NN}) = \langle NN| T(\gamma N\to\pi N)|d\rangle,
\eeq
with the distorted $NN$ wave in the reaction amplitude of the
distorted-wave impulse approximation,
\beq
   T^{\rm DWIA}(E_{NN}) = \langle\psi^{(-)}(NN)| T(\gamma N\to\pi N)|d\rangle.
\eeq
Here we explicitly indicate the energy of the $NN$ state. Also, the
coherent amplitude, with the final bound $NN$ system, is
\beq
   T^{\rm coh}(E_d) = \langle d| T(\gamma N\to\pi N)|d\rangle.
\eeq
Owing to the closure, i.e. a completeness of eigen states of the free $NN$ Hamiltonian 
as well as those of a Hamiltonian with $NN$ interaction,
\beq
 1 = \sum_{NN, \,E_{NN}} |NN\rangle \langle NN| =
  \sum_{NN, \,E_{NN}} |\psi^{(-)}(NN)\rangle \langle\psi^{(-)}(NN)|  ~+~ \sum_d |d\rangle \langle d|,
\eeq
the square of the PWIA off-shell amplitude integrated over all possible
$NN$ states, irrespectively to their energies, exactly coincides with
the square of the DWIA off-shell amplitude  (also integrated over all
possible states) plus the square of the coherent amplitude. In case when
a subset of $NN$ states of certain energies is only considered, as in
the case of finding cross sections at a certain energy, the coincidence
of $|T^{\rm PWIA}|^2$ with $|T^{\rm DWIA}|^2 + |T^{\rm coh}|^2$ is not
strictly valid, however a tendency to have a compensation between the
coherent contribution to the cross section and a decrease in the DWIA
cross section still remains.

\begin{figure}[h!]
\centering\includegraphics[width=0.50\textwidth]{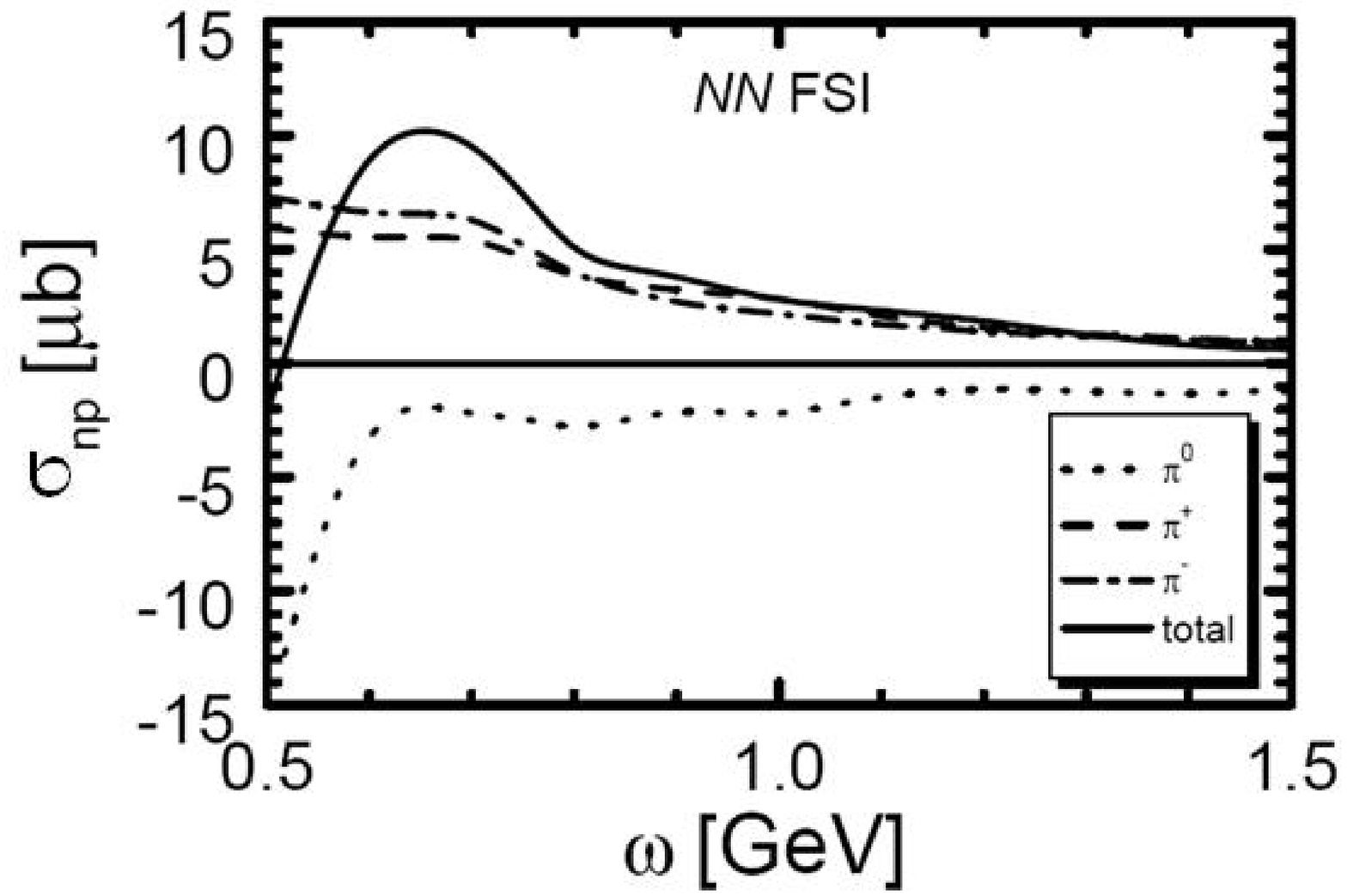}
\includegraphics[width=0.48\textwidth]{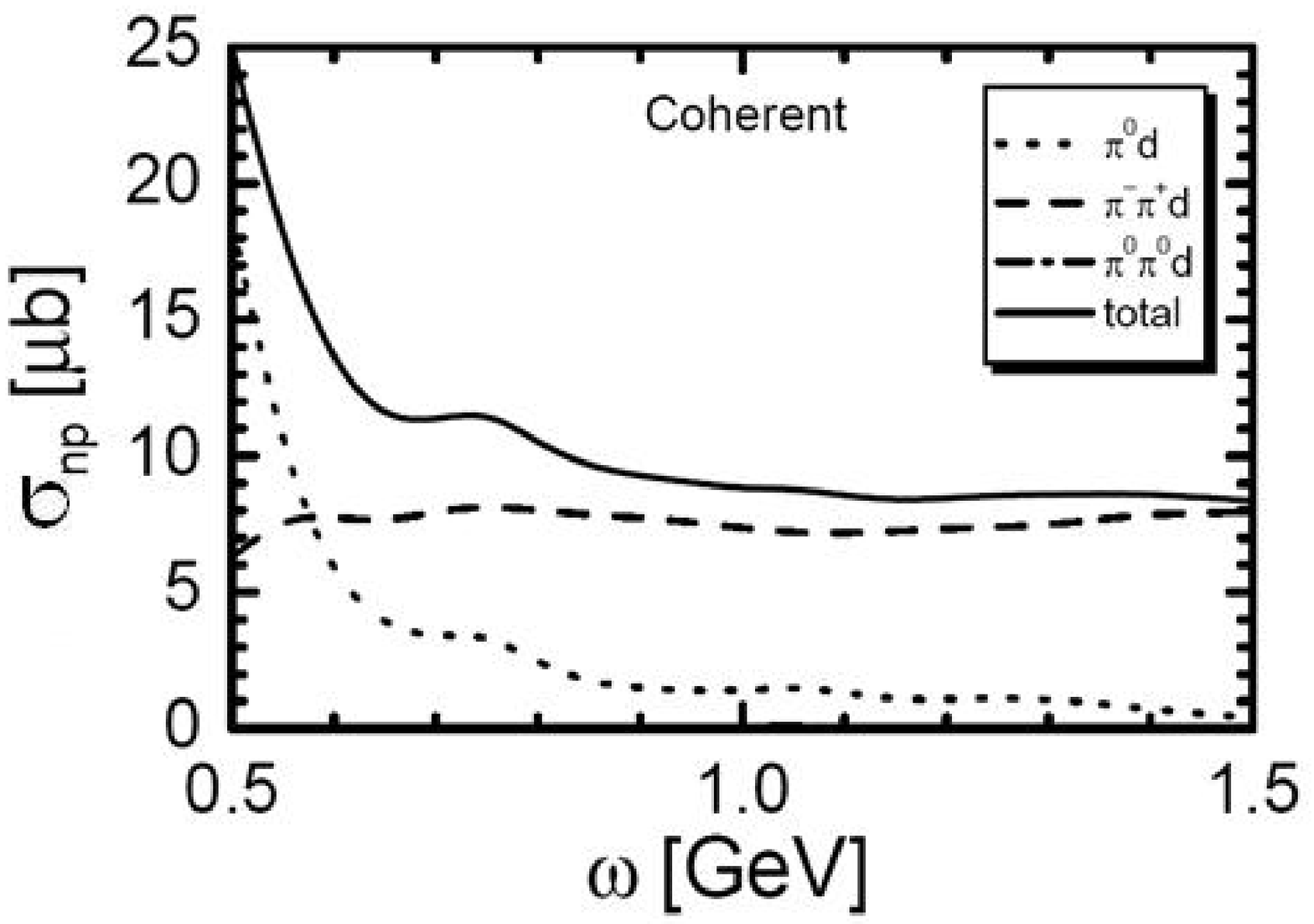}
\caption{Left: Contribution to $\Delta\sigma_{pn}$ due to final state $NN$ interaction in $\gamma d\to\pi NN$.
Right: Contribution to $\Delta\sigma_{pn}$ from $\gamma d\to\pi^0 d$ and $\gamma d\to\pi\pi d$.}
\label{deltaS.NN}
\end{figure}

An illustration of this general tendency can be found in
Fig.~\ref{deltaS.NN} where the negative $NN$-FSI contribution to $\gamma
d\to\pi^0pn$ is close in the magnitude to the positive coherent
contribution to $\gamma d\to\pi^0d$ (see dotted curves).

\subsubsection*{3. $\bm{N\!\!N}$ FSI interaction in 
  $\bm{\gamma d\to\pi\pi N\!\!N}$ and $\bm{\gamma d\to\pi\pi d}$}

\begin{figure}[h!]
\centering\includegraphics[width=0.50\textwidth]{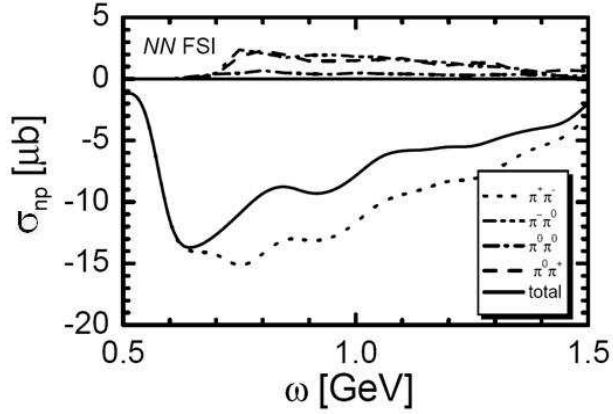}
\caption{Final state $NN$ interaction in $\gamma d\to\pi\pi NN$.}
\label{deltaS.NN2}
\end{figure}

Consideration of the reactions $\gamma d\to\pi\pi NN$ and $\gamma
d\to\pi\pi d$ is similar but more involved owing to a more
complicated structure of the elementary $\gamma N\to\pi\pi N$ amplitude.
We rely here on results obtained by Fix and Arenh\"ovel
\cite{fix05a,fix05b} from which we infer contributions to $\Delta\sigma_{pn}$ shown in
Figs.~\ref{deltaS.NN} (the right panel) and \ref{deltaS.NN2}. Again we see an
essential partial cancelation between $\gamma d\to\pi^+\pi^-d$ and $NN$
-FSI effects in $\gamma d\to\pi^+\pi^-pn$.

\subsubsection*{4. Other small contributions and 
   the net result for $\bm{\Delta\sigma_{\bm{pn}}}$} 

We do not show contributions to $\Delta\sigma_{pn}$ from $\pi N$ FSI in
$\gamma d\to\pi NN$ (found in the described model) and contributions
from the deuteron photodisintegration, $\gamma d \to pn$ (it can be
directly found from experimental data of CLAS \cite{miraz04}) because
they are rather small with the except for energies close to the
$\Delta(1232)$ resonance region. We can anticipate that
$\Delta\sigma_{pn}$ is not affected by $\eta$ meson photoproduction
because $\eta N$ interaction is weaker than that of $\pi N$ and because
effects of $NN$ FSI interaction in the continuum and in the bound state
are again nearly canceled.

\begin{figure}[h!]
\centering\includegraphics[width=0.50\textwidth]{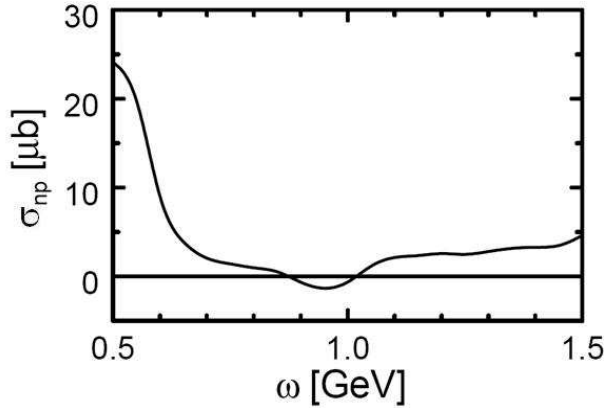}
\caption{Total value of $\Delta\sigma_{pn}$.}
\label{deltaS.tot}
\end{figure}

Taking all contributions together, we arrive at the total value of
$\Delta\sigma_{pn}$ shown in Fig.~\ref{deltaS.tot} which is the main
result of this section. In spite of quite a few pieces of order 10 $\mu$b,
the sum of all contributions to $\Delta\sigma_{pn}$ is found surprisingly small, so that
our improvement to the unfolding procedure is mainly reduced to
a refinement in solving the integral equation.

\subsection*{Extraction of the photoabsorption cross section on the neutron}

Known now all ingredients of Eq.~(\ref{sigma-d}), we can fit
experimental data, determine the unsmeared deuteron cross section
$\sigma_p+\sigma_n$ and then find the neutron cross section
$\sigma_n$. We illustrate this procedure using Daresbury data
\cite{armstrong72a,armstrong72b} for the proton and the deuteron.

\begin{figure}[h!]
\centering
\includegraphics[width=0.49\textwidth]{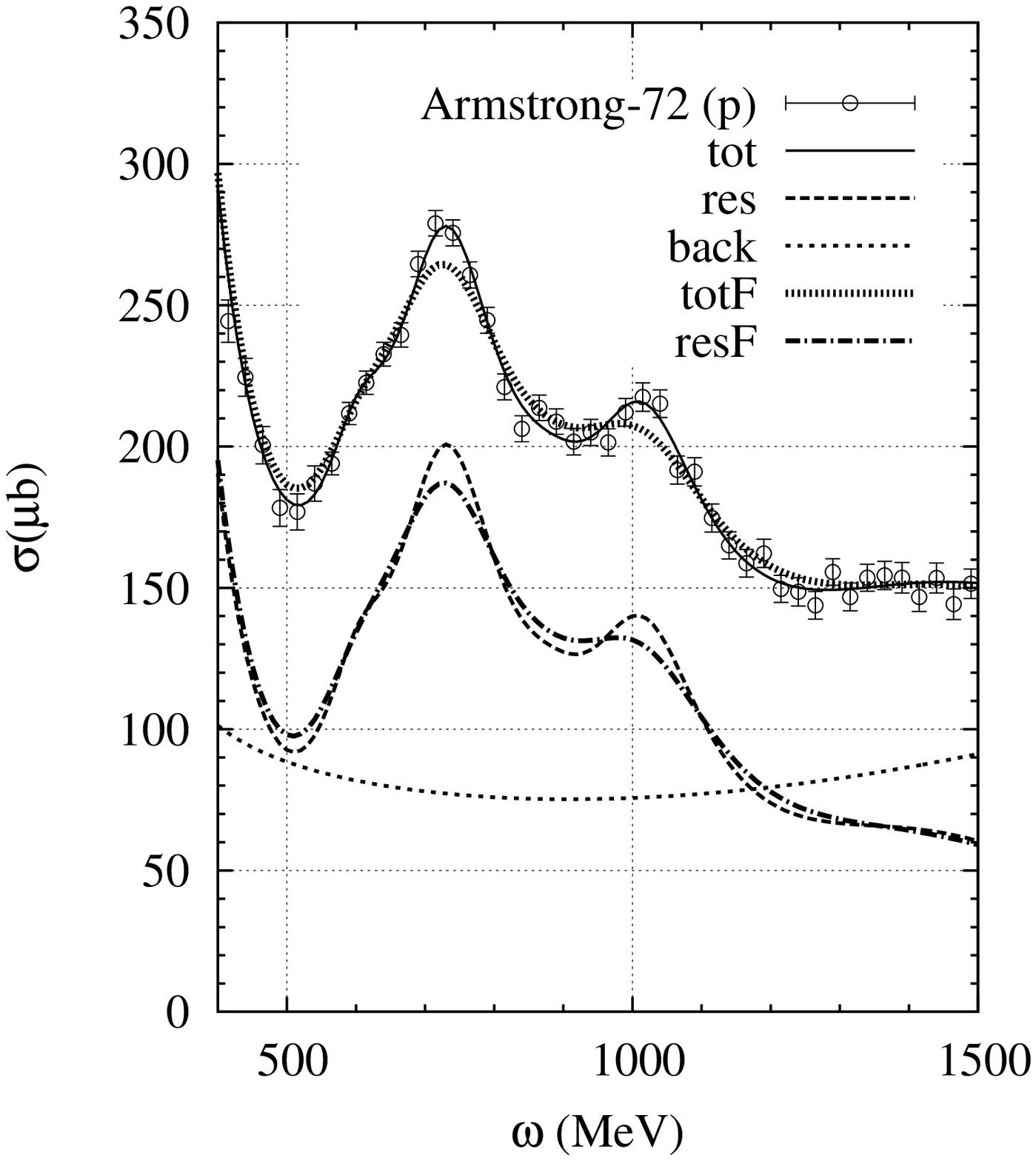}
\includegraphics[width=0.49\textwidth]{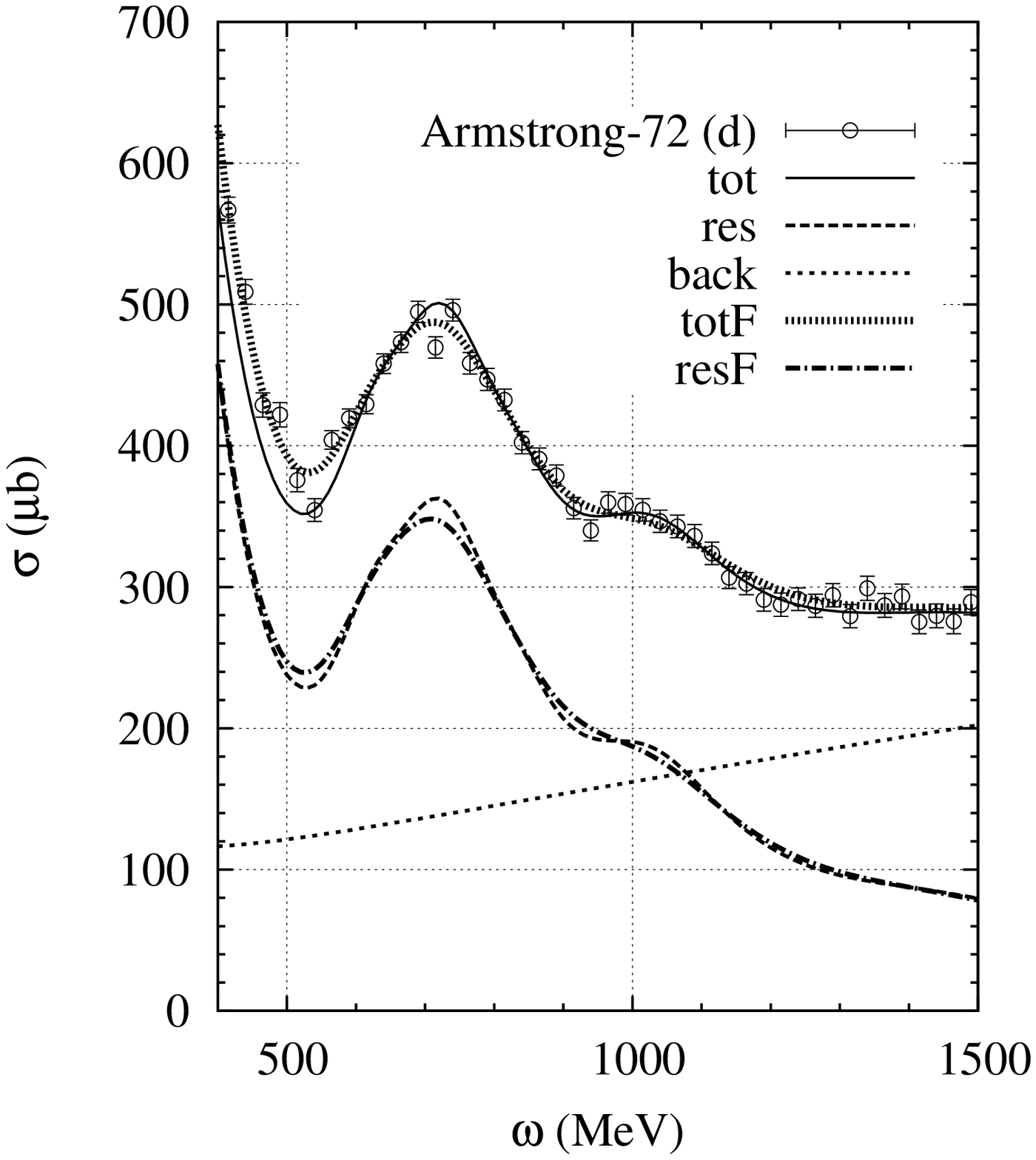}
\caption{Daresbury data for the proton (left) \cite{armstrong72a} and
the deuteron (right) \cite{armstrong72b}, their fit and smearing.}
\label{daresbury}
\end{figure}

Figure \ref{daresbury} (the left panel) shows a smooth fit (the curve
labeled ``tot'') with Eq.~(\ref{sX}) to the experimental proton data and
the result of its smearing with the smearing operator $\hat F$.
Separately shown is the contribution of resonances (and its smearing) and
a smooth background. At the right panel of Fig.~\ref{daresbury} a fitting curve is
shown that, after smearing and adding $\Delta\sigma_{pn}$, comes through
experimental data points (the curve labeled ``totF'').

\begin{figure}[h!]
\centering
\includegraphics[width=0.49\textwidth]{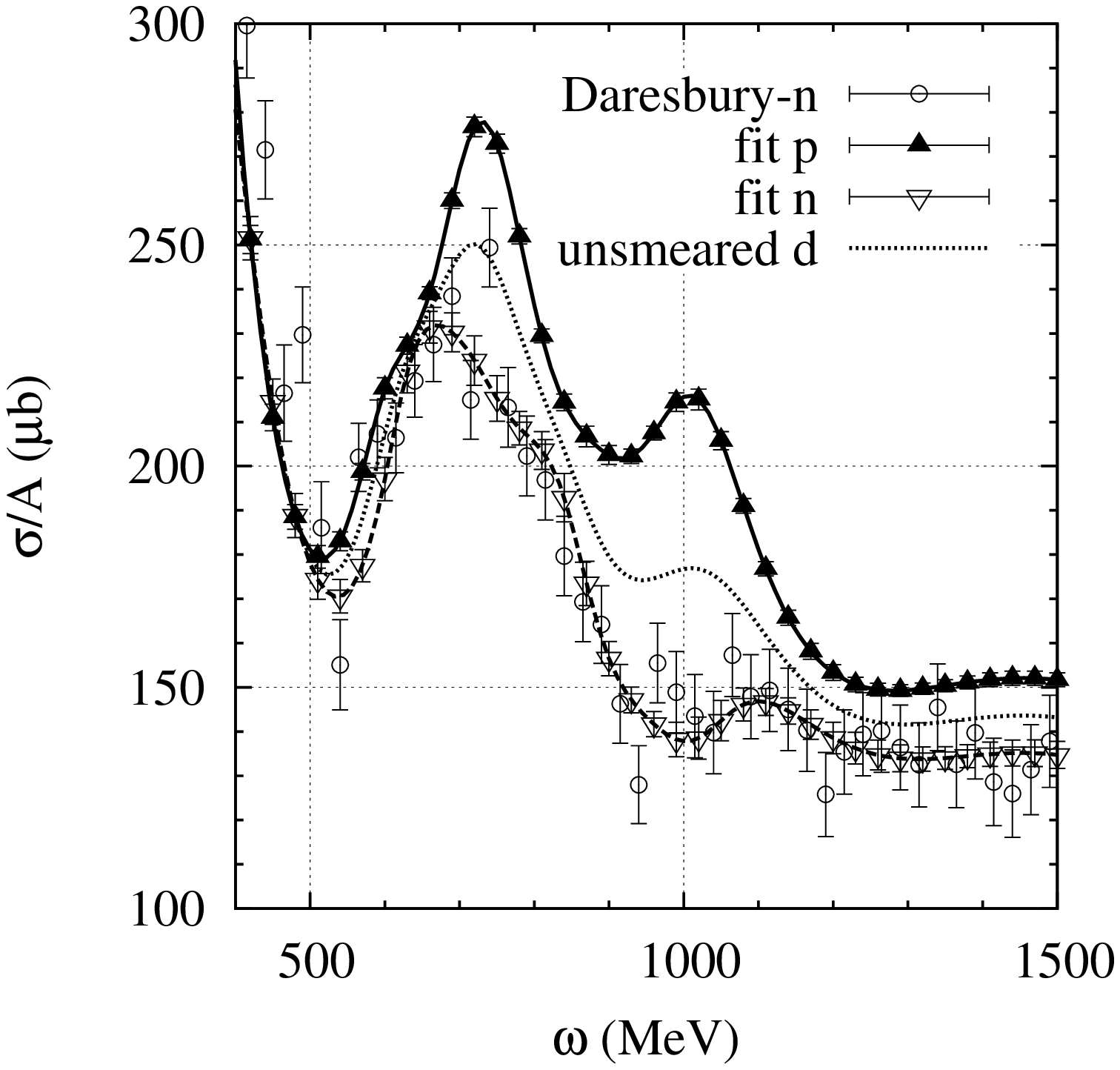}
\includegraphics[width=0.49\textwidth]{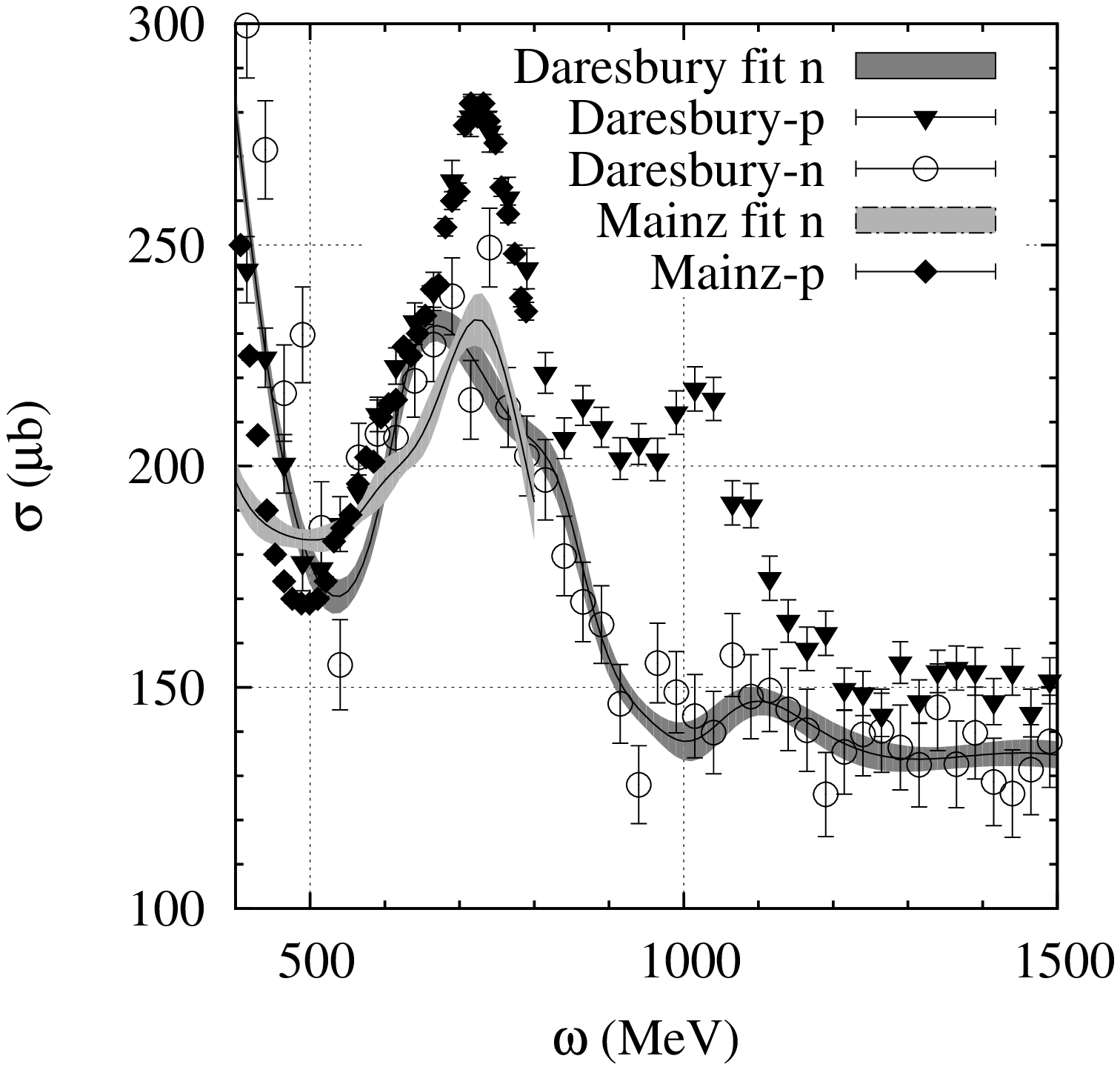}
\caption{Extraction of the neutron cross section $\sigma_n$ from the
deuteron data (Daresbury \cite{armstrong72b} and Mainz \cite{mccormic96}). Original
values of $\sigma_n$ from the Daresbury experiment are shown with open circles.}
\label{daresbury-n}
\end{figure}

From this fit the neutron cross section can be found as a difference,
see Fig.~\ref{daresbury-n} (the left panel). 
In a similar way the neutron cross section can be found from Mainz data \cite{mccormic96}.
Our results are shown in Fig.~\ref{daresbury-n} (the right panel). Bands indicate errors in the found
neutron cross sections there.

\subsection*{Conclusions}

An improved procedure of extracting the total photoabsorption cross
section on the neutron from data on the deuteron is proposed. It
involves a more correct treatment of folding/unfolding of the Fermi
smearing of individual nucleon contributions.

Non-additive corrections are evaluated at medium energies where VMD does
not yet work. They are relatively small in total but they might be more
important in analyses of partial channels of photoabsorption.

We hope that the obtained results will be useful for interpretation of
the GRAAL data and future experiments.

\subsection*{Acknowledgments}

We appreciate very usefull and stimulating discussions with V.G. Nedorezov and A.A. Turinge.

\def\refname{\mbox{\large References}}

\end{document}